\documentclass[%
reprint,
showpacs,
preprintnumbers,
bibnotes,
amsmath,amssymb,
aps,
longbibliography,%
nofootinbib,%
]{revtex4-1}
\usepackage{graphicx}% 
\usepackage{dcolumn}% 
\usepackage{bm}%
\usepackage[
     colorlinks=true,
     linkcolor=black,
     citecolor=black,
	urlcolor=black
]{hyperref}% 
\usepackage{color}
\usepackage{float}
\usepackage{textcomp}
\usepackage{bm}
\usepackage{bbm}
\usepackage{sansmath}
\usepackage{comment}
\usepackage{braket}
\usepackage{ulem}
\usepackage{mathtools}
\usepackage{notes2bib} 

\newcommand{\abohr}{a$_{\text{B}}$}
\newcommand{\mubohr}{\textmu$_{\text{B}}$}

\newcommand{\tx}{\text}
\begin{document}

\title{Quadratic optical response of CrSBr controlled by spin-selective interlayer coupling}

\author{Marie-Christin Hei{\ss}enb\"uttel$^1$}
\email{{m}\_heis08@uni-muenster.de}
\author{Pierre-Maurice Piel$^2$}
\author{Julian Klein$^3$}
\author{Thorsten Deilmann$^1$}
\author{Ursula Wurstbauer$^2$}
\author{Michael Rohlfing$^1$}
\affiliation{$^1$Institute of solid state theory, University of M\"unster, 48149 M\"unster, Germany}
\affiliation{$^2$Institute of Physics, University of M\"unster, 48149 M\"unster, Germany}
\affiliation{$^3$Department of Materials Science and Engineering,  Massachusetts Institute of Technology, Cambridge,  MA 02139, USA}
\date{\today}
 
\begin{abstract}
The optical properties of the layered magnet CrSBr are dominated by
intralayer excitons: the antiferromagnetic order between the layers makes layer-to-layer
charge hopping, and therefore interlayer excitons, spin-forbidden.
An external magnetic field, however, continuously drives the magnetic order towards layer-to-layer ferromagnetic,
which opens spin-allowed charge-transfer channels between the layers.
Here we elaborate how their admixture changes the composition and nature of the excitons, leading to an extension over many layers, and causes a quadratic red-shift
with respect to the external magnetic field.
We address these effects by ab-initio $GW$-BSE calculations as a function of magnetic field and cast the data into a minimal
four-band model to elucidate the interplay between the various interaction and coupling mechanisms.
Our findings should be generally valid for antiferromagnetic layered magnets with and without external magnetic fields,
and moreover for any couple of layers with different spin directions.
Our insights help to systematically address excitons and predict their
optical signatures in such systems.
\end{abstract}

\maketitle

\begin{figure*}[tp]
	\centering  
    \includegraphics[width=2.\columnwidth]{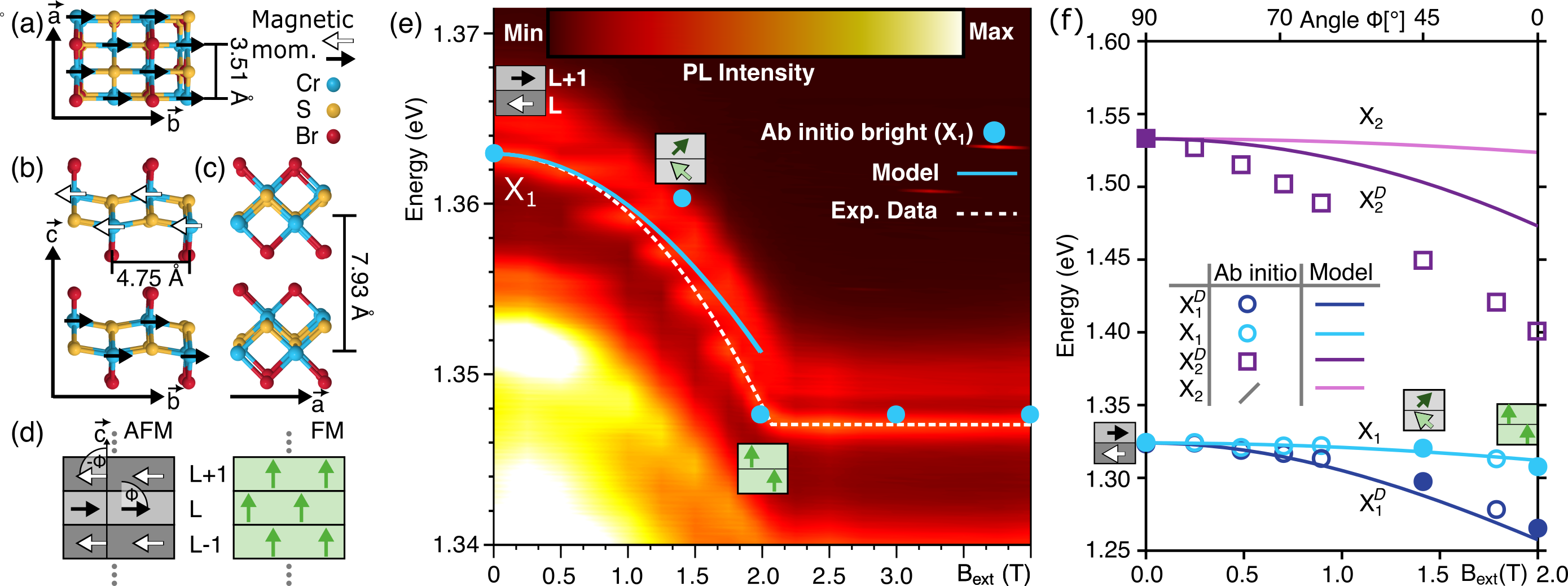}
    \caption{\textbf{Exciton energy shift in external magnetic field.}
    \textbf{a-c} Crystal structure of CrSBr from different viewpoints with arrows denoting the intrinsic magnetization direction. 
    Here, we only show two layers as cutout of the multilayer being a periodic repetition of this scheme. 
    \textbf{d} Magnetic order of the van der Waals coupled layers in the in-plane antiferromagnetic (AFM) state as well as in the out-of-plane ferromagnetic (FM) state with arrows that give the magnetization direction which is uniquely defined by the angle $\phi$ of magnetic moments to the vertical direction. 
	\textbf{e} Photoluminescence (PL) spectra of CrSBr with PL signal  depending on the external magnetic field strength ($B_{\tx{ext}}$). 
    The lowest state from PL that can be attributed to an exciton is marked by the dashed white line, while the broad PL intensity below this state might originate from other excitations like polaritons.  The magnetic state at three different $B_{\tx{ext}}$ is shown as small insets. 
	The blue dots show our exciton energies from ab-initio $GW$-BSE calculations and the blue line denotes our minimal model, both shifted to the experimental data for comparison. 
	\textbf{f} Calculated energies of the lowest two excitons X$_1$, X$^D_1$ and the predominant charge-transfer state X$_2$,X$^D_2$ as a function of the external magnetic field.
    }
\label{fig:fig1}
\end{figure*}

\section*{Introduction.}
Manipulating material properties via external stimuli like screening, 
moiré traps, or magnetism is a key focus in basic research due to its 
easy tunability of opto-electronics~\cite{wang2016screeningExcitonsIOP,Huang2022MoireExcitons,zhaoNatNanotech2017valleySplittingWSe2,Jauregui2019electricControlInterlayer}.
The influence of magnetism, particularly in 2D magnets like CrI$_3$ 
and CrSBr, offers a practical way to control their quantum states~~
\cite{WuLouieNatComm2019cri3Theory,KleinNatPhys2019cri3Interlayer,gibertini2019twoDmagnets}. 
In van der Waals (vdW) layered crystals, such as easily exfoliatable semiconductors that form moiré patterns of two-dimensional (2D) stacked heterostructures~~
\cite{Gong2014InterlayerTransition,Liu2014interlayerTwisting,Huang2022MoireExcitons,Cao2018moireGraphene} or magnetically controlled valley excitons~\cite{Heissenbuettel2021Heterostructure,seylerNanoLett2018}, 
understanding interlayer coupling mechanisms is crucial for further 
development of the intricate field of integrated systems with a variety of applications in ultracompact information technology~\cite{Deng2022NatureUltracompact} or even synaptic devices~\cite{Sun202synapticDevices}. 
Recent work by Wilson et al.~\cite{Wilson2021interlayer} revealed a 
magnetic control mechanism for excitons in antiferromagnetic (AFM) 
bilayer CrSBr, where the magnetization direction affects exciton energies and wavefunctions.
CrSBr, with its semiconducting electronics, large crystal anisotropy, 
and vdW coupled AFM stacked layers, exhibits intriguing phenomena, 
especially air-stability, high Nèel- and Curie-temperatures 
of 132\,K and 160\,K, respectively, or 1D electronics and 
optics~\cite{Goeser1990crsbr,Klein2023CrSBr1D,Lee2021crsbrMagneticOrder,telford2020crsbrMagnetoresistence} 
and is a perfect playground for experiments on magnetism and 
correlated phenomena in two or three dimensions. 
When the magnetic moments are rotated from an AFM coupling 
in-plane to a FM coupling parallel to the external magnetic field,
the lowest exciton in experiments exhibit a quadratic redshift and its 
wavefunction gains significant charge transfer 
(CT)~\cite{Wilson2021interlayer,tabatabavakili2023dopingcontrol}. 
Such a quadratic response is in sharp contrast to
the linear $g$-factors due to the Zeeman shift in common materials \cite{Li_2014,Aivazian_2015,Srivastava_2015,Arora_2016,gfac}.
This provides a manageable technique to control excitons in future 
devices and even switch them to CT excitons with an out-of-plane dipole.
Our experimental and first principles $GW$-BSE data match very 
well, and we are able to develop a general model offering deep 
insights into the magnetic phase dependent physics involved.
The influence of the external magnetic field on CrSBr is expressed 
well in our model and we can reproduce and quantify the redshift of 
exciton energies and increasing CT character.
Despite its simplicity, the proposed model is generally applicable to coupled 2D magnets.
Additionally, our studies unveil that a symmetry forbidden, dark exciton 
is the lowest in energy at finite external magnetic fields.
These findings not only enhance our understanding of analogous 
systems but also open avenues for applications with precise control over switching mechanisms.

\section*{Results.}
We study a multilayer of CrSBr evolving from the layered in-plane AFM order into an external field induced out-of-plane FM order. 
In Fig.~\ref{fig:fig1} (a)-(d) the crystal and magnetic structure of AFM and FM CrSBr is sketched.
Each layer consists of two inner planes of alternating chromium and sulfur atoms embedded in two layers of out-stacking bromine atoms.
The magnetic moments of 3.3\,\mubohr\ from theory are hosted at each chromium atom and align ferromagnetically along the crystal easy axis b. 
A triaxial, anisotropic behavior with hard, middle and easy magnetization axis along c, a, and b, respectively~\cite{Yang2021PRBTriaxialMagnAnisotr} leads to large magnetic, electronic and optical anisotropy, one-dimensional electronics and cigar-like 1D excitons with large binding energy even in bulk material~\cite{Klein2023CrSBr1D}.
Vertically, the monolayers are arranged in AFM A-stacking.
Due to the magnetic order, only the combination of time-reversal and inversion symmetry is preserved in the AFM multilayer leading to the $Pmmn (D_{2h})$ space group. 

An external magnetic field perpendicular to the layers changes the magnetization continuously from AFM into FM order (see Fig.~\ref{fig:fig1} (d)).
For complete rotation from in-plane AFM (along b) to FM (along c),
the external field has to overcome the magnetic anisotropy and AFM coupling energy. 
A clear signature of this changed spin arrangement is given by our
photoluminescence (PL) data at low temperature (4\,K) in Fig.~\ref{fig:fig1} (e), in which
we observe a quadratic redshift of the lowest exciton
marked by the dashed black line starting at 1.362\,eV (AFM) proceeding to 1.347\,eV (FM). 
Note that the broad PL intensity at lower energies
may not be attributed to excitons and originates most likely from self-hybridized polaritons~\cite{WangNature2023CrSBrPolaritons,DirnbergerNature2023CrSBrPolaritons}
or similar, and is therefore not of interest in the scope of this work.

The key issue of our present work is to understand the relationship between external magnetic
field, rotation of the magnetic moments, and its effect on the excitons.
To this end, we first calculate the electronic and excitonic states from first principles ($GW$-BSE calculations in Fig.~\ref{fig:fig1} (f) (dots)) at various intermediate tilting angles $\phi$ (see Fig.~\ref{fig:fig1} (d) for the definition of $\phi$) and then condense the findings and mechanisms into a minimal model.
\begin{figure*}[tp]
	\centering\includegraphics[width=2.\columnwidth]{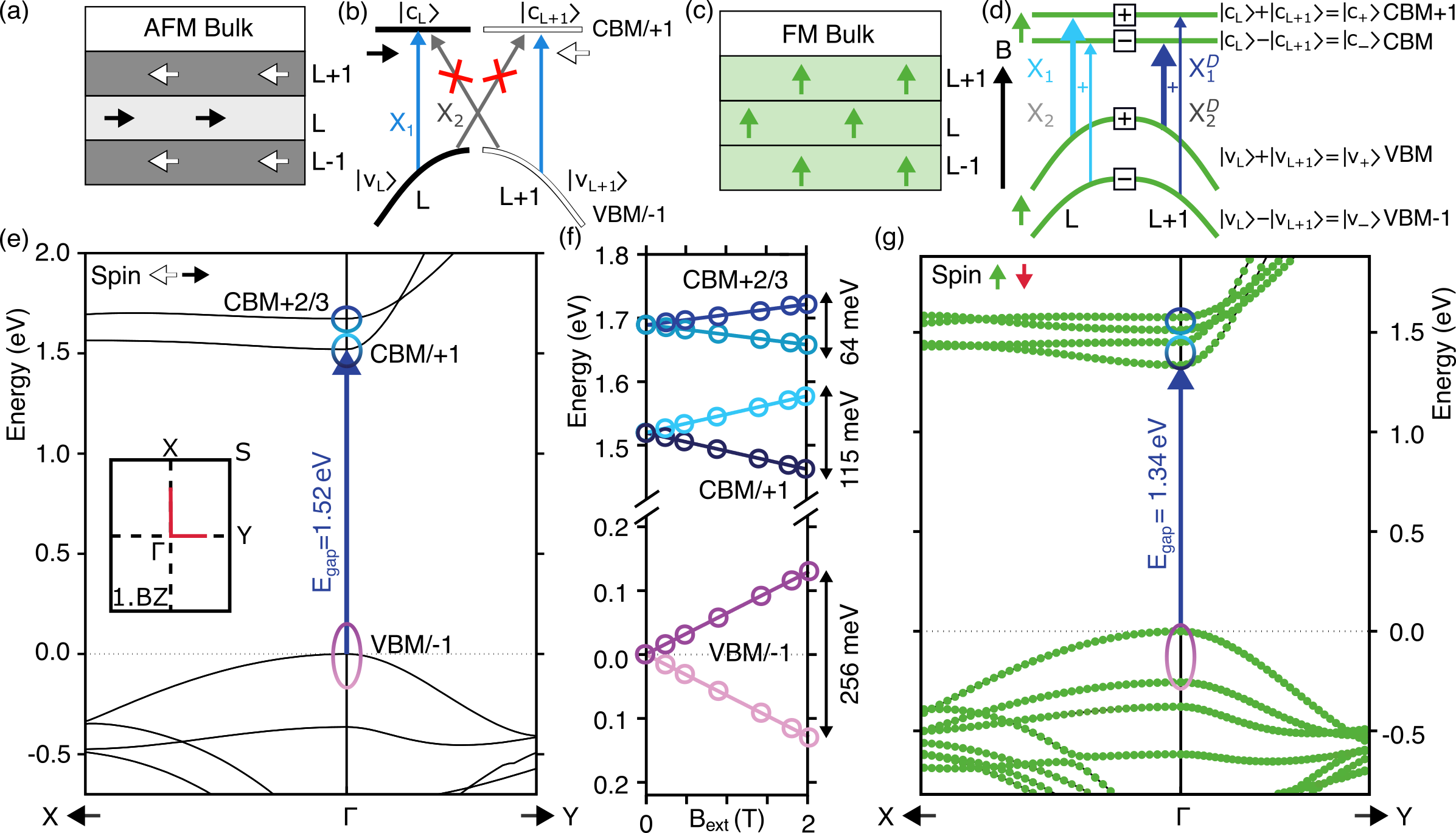}
	\caption{\textbf{Influence of magnetic state on the electronic structure.}
    In \textbf{a} we  show the scheme of the AFM ordered multilayer with arrows giving the magnetization direction of each layer. 
    The sketch in \textbf{b} of the important valence and conduction bands ($\ket{v_{\tx{L,L+1}}}, \ket{c_{\tx{L,L+1}}}$) of layer L/$+$1 shows the possible excitonic transitions (blue arrows) in this magnetic state.
	Only intralayer states (X$_1$) are optically allowed while charge transfer is spin-forbidden (X$_{2}$)
    The same magnetic layer scheme is shown for the FM state in \textbf{c} with splitted energy states in \textbf{d} (VBM/-1, CBM/+1), here,  all in the same spin state (green arrows) aligned parallel to the external field (black arrow) vertical to the monolayer plane. 
    The resulting bands (\textbf{d}) are now linear combinations due to the coupling of all layers (see $\ket{v_{\pm}}\ket{c_{\pm}}$), and the sign in the middle refers to the solutions of our model.  
	The excitons are combined transitions from all layers, so intralayer and charge transfer types are mixed although X$_{1}$/X$^D_{1}$ are still predominantly intralayer excitons. 
    While the lowest transition from the VBM to the CBM is symmetry forbidden, the transition into the CBM+1 is allowed and optically excitable.
    \textbf{e} shows the electronic bandstructure from ab initio $GW$ of multilayer CrSBr in the AFM state around $\Gamma$. 
    Each band is twice degenerate due to the antiferromagnetic order of the layers.  
    The inset shows the 2D part of the Brillouin zone, where the reciprocal direction $\Gamma$-Y corresponds to the real space crystal axis b.
    In \textbf{g} the $GW$ bandstructure of the FM order with colors denoting the spin polarization of each electronic state is shown. 
    Here, all spins are aligned with the external magnetic field and all states split into non-degenerate bands. 
	In between the AFM and the FM bandstructure we depict in \textbf{f} the energy developement of all states from VBM-1 to CBM+3 at $\Gamma$ for an increasing magnetic field and show the  linearly increasing splitting.  }
\label{fig:fig2}
\end{figure*}
Each angle $\phi$ of magnetic moments is attributed to the strength of the external magnetic field between $0$\,T (AFM) up to the saturation magnetization of about $2$\,T (FM).  
The field $B$ gradually forces the moments to align parallel to it ($||$~c) and tilts moments of adjacent layers parallel to each other. 
The magnetization direction in CrSBr is determined by a minimum in total magnetic energy for a given external field.
This is given by three terms: the antiferromagnetic exchange coupling of adjacent layers $E_{J}$, the crystal anisotropy energy $E_{\tx{A}}$, and the Zeeman energy $E_{\tx{Z}}$ that prefers moments parallel aligned to the external field (for details see Supplementary Information (SI) \bibnotemark[Supplement]). 
\bibnotetext[Supplement]{Supplementary Information at \url{link} for details on sample preparation, magnetic switching process and model development as well as on the ab initio data.}
Minimization with respect to the external magnetic field $B$ yields that the 
tilting angle $\phi$ is given by $\cos(\phi) = B/B_{\tx{sat}}$.
At and beyond the saturation field strength $B_{\tx{sat}}$, ferromagnetic order ($\phi = 0^{\circ}$)
is achieved. Our PL data of Fig.~\ref{fig:fig1} (e) show that $B_{\tx{sat}} \approx$ 2.0 T.
For $B < B_{\tx{sat}}$ 
neighboring  layers exhibit alternating orientation of their magnetic moments (L $\hat{=}+\phi$; L$\pm1$ $\hat{=}-\phi$, see Fig.~\ref{fig:fig1} (d).)
In the AFM state ($\phi=\pm 90^{\circ}$) CrSBr consists of two alternating layer types,  L with moments of the majority spin channel along $+$b and L$+1$ along $-$b.
For our discussions only the character of states close to the electronic gap, which are determined by the majority spin channel, are relevant (states of the minority channel are more than $1$\,eV away from the gap).
Moreover, for these states of interest a very small spin orbit coupling (SOC) is found, so that the AFM crystal has entirely decoupled layers with electrons on neighboring layers quasi orthogonal in their spin direction. 

Our calculations find two lowest-energy excitons at about 1.3 eV. In the AFM case ($\phi = 90^{\circ}$) the perfect orthogonality
between the spin channels of neighboring layers causes perfect intralayer nature (i.e., electron and hole reside on the same layer) and degeneracy of the two excitons (at 1.324 eV).
Through the external field and the gradually parallel spin alignment the spin orthogonality is lost and coupling of layers becomes possible.
With increasing $B$ the two excitons split into a bright upper state X$_1$ and an optically forbidden, dark lower state X$^D_1$ (blue dots and lines in Fig.~\ref{fig:fig1} (f)).
They are shifted quadratically down in energy, finally reaching $1.267$\,eV (X$^D_1$) and $1.309$\,eV (X$_1$) in FM order at saturation field strength. 
The state X$_1$ is clearly observed in the measured PL data of Fig.~\ref{fig:fig1} (e), with the same energy downshift of 15 meV from AFM to FM. 
Here, our ab intio data for the redshift agree perfectly with the experiment which is even at quantitative level suprisingly good.
The energetically lower, dark X$^D_1$ with a larger calculated redshift ($57$\,meV) is not observed in the PL data.
We aim to understand and model the detailed physical mechanisms that control all effects resulting from the tuned interaction of layers.  
To keep this model as simple as possible we will consider only transitions between the two highest valence and two lowest conduction bands.

\subsection*{Electronic properties and linear shift.}
To understand the influence of the magnetization direction on the excitons, we have to look firstly at the single-particle physics of the electrons and holes from which the excitons are formed.
Fig.~\ref{fig:fig2} shows the dependence of the electronic states from $GW$ on the magnetic field for the two extreme cases AFM (e), FM (g) and the explicit dependence on the field strength (f). 
For our exciton analysis only the two highest valence and two lowest conduction bands are relevant. 
The AFM order (Fig.~\ref{fig:fig2} (e)) has a direct band gap of $1.52$\,eV at $\Gamma$ with nearly dispersionless bands along $\Gamma$-X resulting from crystal anisotropy as presented before~\cite{Klein2023CrSBr1D}. 
Each line denotes two spin-degenerate states, with each state about 99\% polarized along $+$b/$-$b. 
As mentioned, due to the very small SOC, all of these states stem from the majority spin channel of the two layer types.
In AFM order, the electron wavefunctions of adjacent layers are spin-orthogonal and the electrons cannot tunnel to neighboring layers. 
At finite $B$ the moments are tilted  ($\phi<90^{\circ}$), and the spin states get a component $\sim \cos(\phi) \sim B$ along $\vec{B}$, i.e. perpendicular to the layers (for details see SI).
This causes interlayer interaction and splitting of the formerly degenerate states, proportional to $B$.
Fig.~\ref{fig:fig2} (f) shows our calculated energy shifts at $\Gamma$.  For full FM order (at and beyond $B=2$\, T) we find splittings of $256$ and $115$\,meV for VBM/VBM$-1$ and CBM/CBM$+1$, respectively.
In the saturated FM configuration all bands are spin polarized vertically to the layers, the direct gap reduces to $1.34$\,eV at $\Gamma$ and dispersions along $\Gamma$-X are enhanced (Fig.~\ref{fig:fig2} (g)).
\begin{figure*}[tp]
	\centering\includegraphics[width=2.\columnwidth]{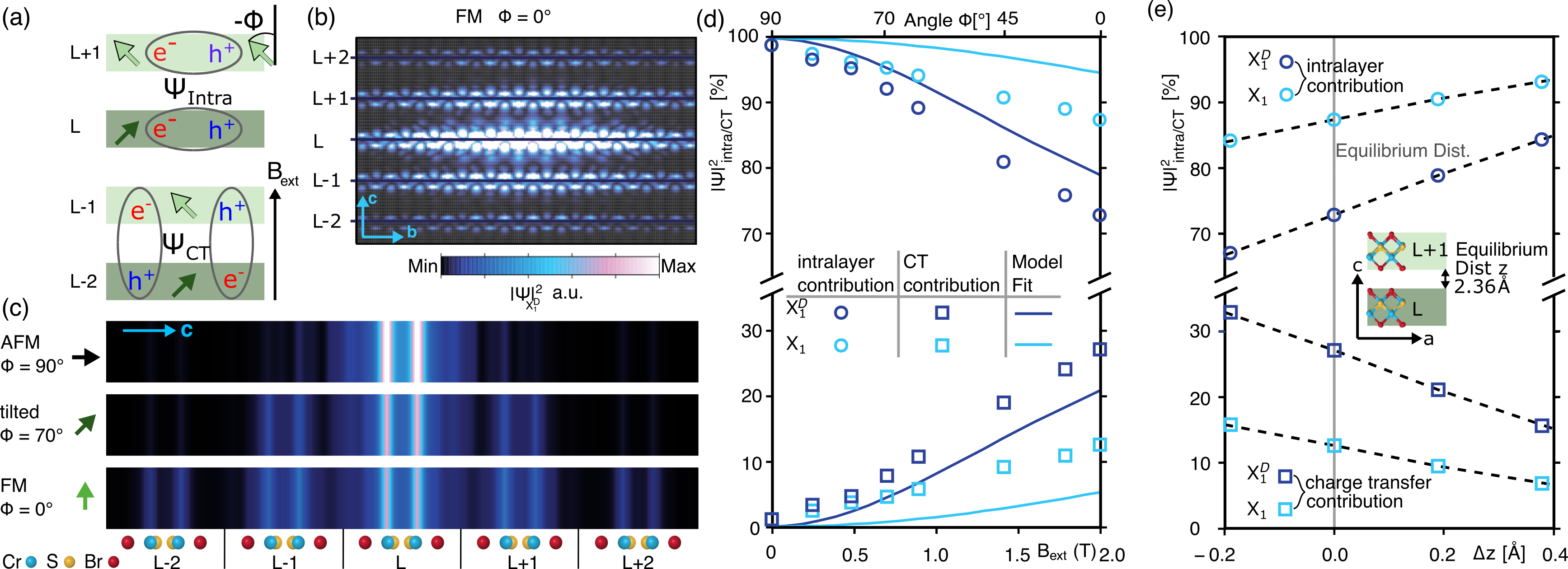}
	\caption{\textbf{Dependency of exciton states on the external field.}
    In \textbf{a} the possible transitions on the layers L and L$\pm$1 or in between are sketched at an arbitrary tilting angle/external field.
    \textbf{b} gives the probability density of the excitonic wavefunction for the first exciton (X$^D_1$) in FM order from our BSE calculation. 
    The wavefunction is projected to the plane vertical to the layers in directions $\vec{b}$ and $\vec{c}$ to show the distribution to the next layers L$\pm$1 and L$\pm$2. 
	Here, the hole is fixed in the center of the plot (middle layer (L)).
	Panel \textbf{c} shows the same quantity along the crystal direction c, after summation over crystal axis b,  for three different magnetic situations.
    The electron probability density strongly depends on the magnetic state which is given by the angle and arrow aside the plot. 
	\textbf{d} shows our $B_{\tx{ext}}$ dependent ab initio results (dots) and model functions (lines) of the intralayer and charge-transfer components for X$^D_1$ and X$_1$.
    Also in \textbf{e} the intralayer and charge transfer contributions to the wavefunctions of X$_{1}$/X$^D_{1}$ are given, here with dependence on the vertical distance $\Delta z$ seperating layers L and L$+1$. The dashed lines are guide to the eye. }
\label{fig:fig3}
\end{figure*} 

A simple model allows to describe the influence of the layer coupling on the electrons and thus reproduces our $GW$ results.
We introduce a coupling Hamiltonian in the electronic basis states on layer L and L$+1$ (see Fig.~\ref{fig:fig2} (b)).
The coupling strength is proportional to the parallel aligned spin component $\cos(\phi)$ (i.e. $\sim B$ , for details see SI) while the proportionality constant is given by the finite overlap of the spatial wavefunctions.
For each pair of degenerate bands this leads to a $2\times 2$ coupling Hamiltonian
\begin{equation}\label{eq:Hel}
  H_{\bm{k}}^{\tx{el}}(B) =
    \begin{pmatrix}
      E_{n \bm{k}} & t_{n \bm{k}}  B\\
      t_{n \bm{k}}  B & E_{n \bm{k}}
    \end{pmatrix} 
    \qquad 
    \begin{matrix}
	    (\leftrightarrow\ket{n_{\tx{L}}~~~\bm{k}})\\
		(\leftrightarrow\ket{n_{\tx{L+1}}\bm{k}})
    \end{matrix}
  \qquad .
\end{equation}
From the bandstructure (Fig.~\ref{fig:fig2} (g)) we find that all quantities are $\bm{k}$ dependent. 
Here, $E_{n \bm{k}}$ is the degenerate eigenenergy of state $\ket{n \bm{k}}$ in the AFM order and $t_{n\bm{k}}$ is the coupling strength or tunneling parameter of a particle between L and L$+1$.
The diagonalization leads to eigenenergies $E_{n \bm{k},\pm} = E_{n \bm{k}} \pm t_{n \bm{k}} B$ and wavefunctions which are the corresponding positive and negative linear combinations of bands from each single layer, denoted in Fig.~\ref{fig:fig2} (d) e.g. $\ket{v_{\pm}}$ (details see SI).
From the linear fit of our ab initio data at $\Gamma$ in Fig.~\ref{fig:fig2} (f) we get the parameters $t_{n \bm{k}}$ of the VBM $t_{v, \Gamma}= 64\,$meV/T and the CBM $t_{c, \Gamma}= 29\,$meV/T for a hole/electron at $\Gamma$, respectively.

\subsection*{Optical properties and quadratic shift.}
Having understood the coupling of electronic states, we get back to the behavior of excitons in Figs.~\ref{fig:fig1} (e) and (f).
In the absence of interlayer coupling (AFM) the first degenerate exciton pair is at $1.324$\,eV.
We can describe these states in two different ways.
Either we explain them as transitions of an electron from the VBM to the CBM on each layer, L or L+1 seperately (Fig.~\ref{fig:fig2} (b)),
or we use the combined basis of $\ket{v_{\pm}}$ and $\ket{c_{\pm}}$ (see Fig.~\ref{fig:fig2} (d) with zero splitting),
which is convenient for $B>0$ but also for $B=0$\,T. 
Four transitions $\ket{v_+c_+}$, $\ket{v_-c_-}$, $\ket{v_+c_-}$, and $\ket{v_+c_-}$) are possible between these states. 
Due to the  symmetry of the system (i.e. $PT$~\cite{Klein2023CrSBr1D} the first two and last two transitions do not mix (see SI).
Here, we focus on $\ket{v_+c_+}$ and $\ket{v_-c_-}$ which results in a $2\times 2$ Hamiltonian for the bright excitons.
The other two transitions can be treated analogously.
\begin{align}\label{eq:HBSE}
   &H^{\tx{BSE}} = \nonumber\\
    &\begin{pmatrix}
      E_{\tx{g}} + t_{e+h} B + V_{eh} & \Delta V_{eh}\\
      \Delta V_{eh}  & E_{\tx{g}} - t_{e+h} B + V_{eh}
    \end{pmatrix}~~~
   	 \begin{matrix}
		 (\leftrightarrow\ket{v_{+}\,c_{+}}) \\ 
		 (\leftrightarrow\ket{v_{-}\,c_{-}})
   	 \end{matrix}
\end{align}
Neglecting the electron-hole interaction $V_{eh}$ and $\Delta V_{eh}$ 
the diagonal would describe uncorrelated interband transitions between the bands, 
i.e. the gap energy $E_\text{g}$ and its shift by the combined tunneling 
parameters of electron and hole $t_{e+h} = t_h + t_e$.
The electron-hole interaction $V_{eh} = (V_{\tx{IN}} + V_{\tx{CT}})/2$ enters on the diagonal,
while the off-diagonal is given by the difference of intralayer ($V_{\tx{IN}}$) and interlayer ($V_{\tx{CT}}$) interaction
$\Delta V_{eh} = (V_\tx{IN} - V_\tx{CT})/2$ for electron and hole on the same (IN) and different layers (CT), respectively.
We underline that only the energy splitting of the bands changes with $B$,
whereas the electron-hole interaction is constant.
Nevertheless, its effect on the excitonic states varies with $B$ and results in changes of the exciton energy and its character as we show in our analysis of states below.
From the previous considerations in Eq.~(\ref{eq:Hel}) on the basis of ab-initio $GW$ 
we use $t_{h} = -0.064\,\tx{eV/T}\cos(k_z c)\delta_{k_z,k'_z}$
and $t_{e} = 0.029\,\tx{eV/T}\cos(k_z c)\delta_{k_z,k'_z}$
including the $k_z$-dependence of $N_{k_z}$ points in the crystal $c$ direction,  with $N_{k_z}$ being the number of $k_z$ grid points in that direction to be employed in the BSE calculation for the excitons.
From our ab-initio BSE we find $V_{\tx{IN}} = -0.350\,\tx{eV}/N_{k_z}$ and $V^{\tx{CT}} = -0.141\cos((k_z-k'_z) c)\,\tx{eV}/N_{k_z}$.

The diagonalization of Eq.~(\ref{eq:HBSE}) yields the dominant behavior $\sim B^2$ of the exciton energies
and mixes the corresponding transitions $\ket{v_+c_+}$ and $\ket{v_-c_-}$ to X$_{1,2}$.
From our ab initio data we find X$_{2}$ at $1.533$\,eV in the AFM state shifting to $1.401$\,eV in the FM state (see Fig.~\ref{fig:fig1} (f)).
For the symmetry forbidden, dark transitions ($\ket{v_+c_-}$,$\ket{v_+c_-}$) the Hamiltonian $H^\tx{BSE}_D$ looks similar to Eq.~(\ref{eq:HBSE}) with one important distinction:
Instead of $t_{e+h}$ the difference of tunneling $t_{e+h,D} = t_h - t_e$ enters.
The corresponding excitons X$^D_{1}$ and X$^D_{2}$ are lower in energy compared to their bright counterparts (see Fig.~\ref{fig:fig1} (f)).
With increasing $B$, X$^D_{1}$ and X$_{1}$ split due to the different tunnelings and decreasing gaps  of $\ket{v_+c_-}$ and $\ket{v_+c_+}$.
Considering the simplicity of the model, the agreement with the ab-initio $GW$/BSE results
(circles/squares in Fig.~\ref{fig:fig1} (f)) is very good, especially for the lowest states.
The model fully explains the experimentally measured redshift in Fig.~\ref{fig:fig1} (b)) of X$_{1}$.\\
The nature of the excitons is governed by the relative alignment of the spins
which leads to the different tunnelings $t_h$ and $t_e$.
For a better understanding, we analyse their character in the following.

\subsection*{Controlling the charge transfer contribution.}
The basis of $\ket{v_{\pm}}$ and $\ket{c_{\pm}}$ is symmetric on both layers,  L and L+1, and may hide the localization of each exciton at first glance.
Thus,  in Fig.~\ref{fig:fig3} we analyze the internal structure of the excitons. 
Panels (b) and (c) show the probability distribution of the electron relative to the hole (located on the middle layer), for the state X$^D_1$  (the state X$_{1}$ looks similar).
The lowest excitons of CrSBr are influenced by the large crystal anisotropy and the flat bands along $\Gamma$-X lead to extended 1D like exciton wavefunctions along axis b (Fig.~\ref{fig:fig3} (b)). 
When the external magnetic field is turned on, the enabled layer interaction induces hopping of the electron from layer L (position of the hole) to layer L$\pm$1 and farther. 
In Fig.~\ref{fig:fig3} (c) we show the projection of the exciton wavefunction onto the axis c,  evolving from being restricted to one layer to becoming more extended to the adjacent layers with increasing tilting angle.
In the AFM state the exciton is fixed to layer L and its character is intralayer while in the FM state the exciton is extended and gets significant charge transfer contributions. 
For the analysis of this CT of X$^D_{1}$ and X$_{1}$ we sum up the probability density on the layer with the hole (intralayer) and the probability on the adjacent layers (CT) and give the percentages in Fig.~\ref{fig:fig3} (d). 
The ab initio results (dots) at $B=0$\,T show decoupled layers, where both excitons are vertically restricted to the respective layer (99\% intralayer character). 
With increasing magnetic field we find that the states get a significant CT with up to 27\% (X$^D_1$) and 13\% (X$_1$) in the FM order. 

From our minimal model the contributions can be evaluated as sum and difference of the two evolved transitions.
E.g. for $B=0$ the contributions of $\ket{v_+c_+}$ and $\ket{v_-c_-}$ are identical,
their sum (intralayer contribution) is 1, their difference (CT contribution) is zero.
Qualitatively we find the same behavior with increasing $B$ as in our ab initio results shown as the blue curves in Fig.~\ref{fig:fig3} (d). 
To first non-vanishing order in $B$ the model yields the CT contribution $\sim t_{e+h}^2 B^2$
and explains the smaller contribution of X$_1$ ($t_{e+h}^2 \sim 0.035^2$) compared to X$^D_1$ ($t^2_{e+h,D} \sim 0.093^2$).
In saturation we find 21\% (X$^D_1$) and 5\% (X$_1$) CT which is similar to the ab initio results discussed above.
Thus the energetically lowest excitons remain intralayer type with significant CT contributions at finite fields. 
The higher-energy excitons X$^D_2$ and X$_2$ (1.533\,eV in AFM order) show the opposite behavior, starting at 100\% CT character and gaining some intralayer share towards FM.

This analysis fully explains the increasing CT admixture of $X_1$ and $X^D_1$ observed in our ab-initio data. Quantitative agreement, however, is not achieved (see e.g. Fig.~\ref{fig:fig3} (d)) due to the minimal nature of our model (e.g., the number of bands included).

These findings suggest further ways to manipulate the energy and CT character in CrSBr. 
With a variation of the layer distance we can change the overlap of orbitals (mostly given by Br atoms) that mediate the interlayer interaction and thus the size of tunneling $t_{e/h}$. 
In Fig.~\ref{fig:fig3} (e) we show the dependence of the CT contribution to X$^D_{1}$/X$_{1}$ on the relative variation $\Delta z$ of layer distance with respect to the equilibrium position.
If the distance is compressed by 0.2\, \AA,~the CT contribution is increased from 27\% to 33\%, while a lattice expansion by 0.2\, \AA~leads to 21\% CT contribution to the X$^D_1$ (for details see SI).
In experiment the interlayer distance may be controlled e.g. by pressure as done in \cite{PawbakeMagnetoSensing2023CrSBr} and our findings can explain excitonic behavior under such conditions.\\

In summary, we explain the underlying mechanisms governing the behavior of electrons and excitons in multilayer CrSBr under the influence of an external magnetic field. 
Our ab initio $GW$-BSE calculations accurately reproduce the redshift of the lowest visible exciton and reveal a splitting of formerly spin-degenerate states, proportional to the magnetic field.
The energy redshifts,  CT admixture, and band gap reduction are explained well in our simple but expressive model set up by parameters from the ab-initio data. 
External magnetic fields enable the layer coupling and induce a substantial CT character of the excitons of up to 27\% in the ferromagnetic order, which is further manipulable by adjusting the interlayer distance.
Of particular interest is the identification of a dark symmetry-forbidden exciton state initially degenerate with the allowed second exciton in the antiferromagnetic crystal. 
This state, shifts to energies below the visible bright exciton observed in photoluminescence at finite fields.
Our findings pave the way for a systematic control of electrons, excitons, charge transfer dynamics, vertical extension of wavefunctions, and binding energies through the application of external magnetic fields.

\section*{METHODS}
\subsection*{Experimental characterization and photoluminescence spectroscopy}
Bulk-like multilayer samples of CrSBr have been prepared by micromechanical cleavage from bulk crystals onto SiO$_2$/Si substrates using Nitto tape. 
The thickness of $70$\,nm and $225$\,nm of the multilayer flakes has been determined by atomic force microscopy in ambient using a XE 100 from Park Systems with a cantilever of the NSC 15 series in the non-contact mode.
The samples are characterized by non-resonant Raman spectroscopy using linearly polarized light from a 532\,nm (2.33\,eV) wavelength laser diode. 
In order to determine the in-plane crystallographic a and b directions that are highly anisotropic, the linearly polarized light in the excitation path has been rotated with respect to the crystal axes and the co-linear polarized Raman response has been determined. 
The intensity of first order Raman modes are sensitive to the crystallographic direction~\cite{Klein2023CrSBr1D}. 
Our magneto-photoluminescence measurements were conducted at 4\,K in Faraday configuration utilizing a pulse tube refrigeration operating with helium gas (Bluefors) equipped with windows for optical access and with a superconducting 7\,T electromagnet. 
The sample was mounted on a cold finger equipped with $x-y-z$ piezo stages for sub-micrometer positioning accuracy. 
The laser was focused on the sample in back-scattering geometry with a low temperature compatible objective lense with a numerical aperture of 0.82 mounted on a homebuilt stage attached to the 4\,K stage of the cryostat.  
The focused spot on the sample had a diameter of $\approx 2$\,µm in the relevant wavelength range. 
The sample was excited close to resonance at a wavelength of 850\,nm (1.46\,eV) using a continuously frequency tunable continuous wave (CW) Ti:sapphire laser. 
The emitted light was guided to the entrance slit of a spectrometer with 750\,nm focal length and dispersed using a 600 lines/mm grating resulting in a spectral resolution of 0.1\,nm within the relevant wavelength range of 900\,nm to 950\,nm. 

\subsection*{Many body perturbation theory from the $GW$ and BSE scheme}
To simulate the experimental samples we study bulk CrSBr.
The description requires arbitrary spin configurations,
i.e.  in-plane ferromagnetic order, antiferromagnetic order from adjacent layers, and different tilted spin configuration in between.
We cannot neglect spin-orbit coupling effects although the effects for the states close to the gap are small.
For energetically higher and lower states, as well as for the correct description of symmetry in this system, spin-orbit coupling is taken fully into account. 
Therefore, we calculated the electronic ground state from first principles by employing density functional theory (DFT) in the noncollinear formalism of the spin density approach \cite{stark2011magnetic,sandratskii1998noncollinear}.
For the exchange correlation functional we applied the generalized gradient approximation (GGA) \cite{Perdew1992gga}.
The resulting energies and spinor wavefunctions are used as a starting point for many body perturbation theory including the noncollinear magnetism and spin-orbit coupling effects. 
In the $GW$ approximation \cite{hedin1965new} we calculated the self-energy $\Sigma=iGW$ from the one-particle Green's function $G$ and the screened Coulomb interaction $W$ (including the dielectric response in random phase approximation) which is then used to calculate the quasiparticle (QP) bandstructures. Here, $\Sigma$ replaces the DFT exchange correlation energy $V_{\text{xc}}$ and leads to the QP Hamiltonian and eigenenergies
\begin{align}
    H^{\text{QP}} &= H^{\text{DFT}} + iGW - V_{\text{xc}} \\
    E_{n\mathbf{k}}^{\text{QP}} &= E_{n\mathbf{k}}^{\text{DFT}} + \bra{\psi_{n\mathbf{k}}^{\text{DFT}}} \Sigma\left( E^{\text{QP}}_{n\mathbf{k}}\right)-V_{\text{xc}}\ket{\psi_{n\mathbf{k}}^{\text{DFT}}} ~ .
\end{align}
While we do not extend our DFT description by adding an arbitrary Hubbard $U$, we note the related physical effects like on-site Coulomb repulsion and Hubbard correlation are included in the self-energy.
In this QP approach we employ the so-called one-shot $GW$ method including non-diagonal terms of the self-energy as we have previously found that the diagonal parts are not sufficient for CrSBr~\cite{deilmann2020valley,Heissenbuettel2021Heterostructure,forster2015two}.
The solution of the full eigenvalue problem yields the coefficients $D^n_{m\mathbf{k}}$ for the linear combination of DFT states to the new QP wavefunctions (for details see Ref.~\cite{Heissenbuettel2021Heterostructure})
\begin{align}
    \sum_{n'} H^{\text{QP}}_{nn'}(\mathbf{k};E^{\text{QP}}_{m\mathbf{k}}) D^{n'}_{m\mathbf{k}} &= E^{\text{QP}}_{m\mathbf{k}} D^n_{m\mathbf{k}} \\
    \psi_{n\mathbf{k}}^{\text{QP}} &= \sum D^n_{m\mathbf{k}} \psi_{n\mathbf{k}}^{\text{DFT}} ~ .
\end{align}
To anticipate the large opening of the band gap from DFT to QP bandstructure and to circumvent a full self-consistent $GW$ to determine the shifts we applied a scissors operator of 1.0\,eV.\\
To evaluate the optical behavior of CrSBr including the electron-hole interaction we solve the Bethe-Salpeter equation (BSE)~\cite{RohlLouiePRB2000} on top of the $GW$ approximation. For exciton state $S$ this is given by 
\begin{equation}
    (E_{c\bm{k}} - E_{v\bm{k}})A^S_{vc\bm{k}} + \sum_{v'c'\bm{k'}} K_{vc\bm{k},v'c'\bm{k'}}(\Omega^S)A^S_{v'c'\bm{k'}} = \Omega^S A^S_{vck} \qquad ,
\end{equation}
with the interaction kernal $K_{vc\bm{k},v'c'\bm{k'}}$ in Tamm-Dancoff approximation, the energy $\Omega^S$ of exciton state $S$ and the amplitudes $A^S_{vc\bm{k}}$ that set up the exciton wavefunctions 
\begin{equation}
    \Psi_S(x_e,x_h) = \sum_{vc\bm{k}} A^S_{vc\bm{k}} \Psi^{\dag}_{v\bm{k}}(x_h) \Psi_{c\bm{k}}(x_e) \qquad .
\end{equation}
Here, $x_{e/h} = (\bm{r}_{e/h}, \sigma_{e/h})$ is the combination of position and spin and the results from Fig.~\ref{fig:fig3} correspond to $\sum_{vc\bm{k}} |A^S_{vc\bm{k}}|^2$.  
We include the off-diagonal elements of the self-energy as the BSE is solved in the original DFT basis as explained in details in Ref.~\cite{Heissenbuettel2021Heterostructure}.\\
For our quasiparticle calculations we employ a hybrid basis set of Gaussian orbitals with decay constants from 0.14 to 12.5\,\abohr$^{-2}$ combined with plane waves with an energy cutoff set to 1.5\,Ry
to represent of all two-point functions ($P,\varepsilon,W$) in $GW$. 
Moreover, we apply a $k$-point sampling of $30\times 22 \times 4$ points in the first Brillouin zone for well converged $GW$ results and use, due to the numerical costs for the BSE, a grid of $16 \times 12 \times 4$ points, which is sufficiently converged for the lowest exciton states.

\section*{AUTHOR INFORMATION}
\subsection*{Corresponding Author}
\textbf{Marie-Christin Heißenbüttel} -- \textit{Institute of solid state theory, University of M\"unster, 48149 M\"unster, Germany}

\subsection*{Authors}
\textbf{Pierre Piel} -- \textit{Physical institute, University of M\"unster, 48149 M\"unster, Germany}

\textbf{Julian Klein} -- \textit{Department of Materials Science and Engineering,  Massachusetts Institute of Technology, Cambridge,  MA 02139, USA}

\textbf{Thorsten Deilmann} -- \textit{Institute of solid state theory, University of M\"unster, 48149 M\"unster, Germany}

\textbf{Ursula Wurstbauer}  -- \textit{Physical institute, University of M\"unster, 48149 M\"unster, Germany}

\textbf{Michael Rohlfing} -- \textit{Institute of solid state theory, University of M\"unster, 48149 M\"unster, Germany}

\subsection*{Notes}
The authors declare no competing financial interest.

\begin{acknowledgements}
The authors gratefully acknowledge the financial support
from German Research Foundation (DFG Project No. DE 2749/2-1 and No. WU637/4-2, 7-1), the DFG Collaborative Research Center SFB 1083 (Project No. A13), and computing time granted by the John von Neumann Institute for Computing (NIC) and provided on the super-computer JUWELS at Jülich Supercomputing Centre (JSC).
\end{acknowledgements}

%%bib
\section*{REFERENCES}
\bibliography{paper}

%\vspace{2cm}
%\begin{Large}
%\textbf{Graphical TOC Entry}
%\end{Large}
%\begin{figure}[h!]
%\centering
%\includegraphics[width=8.76cm]{toc.pdf}
%\caption{For TOC only}
%\end{figure}

\end{document}